%August 1996
\input harvmac
\noblackbox
\Title{\vbox{\baselineskip12pt\hbox{UCLA/96/TEP/26}\hbox{hep-th/9608158}}}
{\vbox{\centerline{A Relation between
the Anomalous Dimensions}\vskip2pt\centerline{and OPE Coefficients
in Asymptotic}
\vskip2pt\centerline{Free Field Theories$^\star$}}}
\footnote{}{$^\star$ This work was supported in part by the U.S. Department
of Energy, under Contract DE-AT03-88ER 40384 Mod A006 Task C.}

\centerline{Hidenori SONODA$^\dagger$\footnote{}{$^\dagger$
E-mail: sonoda@physics.ucla.edu} and Wang-Chang SU$^*$\footnote{}{$^*$
E-mail: suw@physics.ucla.edu, Address after 1 October 1996: Department of
Physics, National Tsing-Hua University, Hsinchu, Taiwan}}
\bigskip\centerline{\it Department of Physics,
UCLA, Los Angeles, CA 90095--1547, USA}

\vskip 1in
In asymptotic free field theories we show that part of the OPE
of the trace of the stress-energy tensor and
an arbitrary composite field is determined by the anomalous dimension
of the composite field.  We take examples from the two-dimensional
O(N) non-linear sigma model.

%\draft
\Date{August 1996}

%begin definitions
\def\no{\noindent}

\def\dt{{d \over dt}}
\def\vev#1{\left\langle #1 \right\rangle}

\def\H{\Theta}
\def\ep{\epsilon}
\def\dO{d\Omega}
\def\C{{\cal C}}
%end definitions

It is well known that in a renormalizable field theory the anomalous dimensions
of composite fields determine the leading behavior of their OPE (operator
product expansion) coefficients \ref\rtext{For example, D.~J.~Gross, {\it in}
Methods in Field Theory, eds. R.~Balian and J.~Zinn-Justin (North-Holland
and World Scientific, 1981)}.
In this note we wish to show that part of the OPE
of the stress-energy tensor and an arbitrary composite field
is determined by the anomalous dimension of the composite field.
Let us recall that in two dimensional
conformal field theory the OPE of the stress-energy tensor
and a conformal field is completely determined by the scale dimension of the
conformal field \ref\rbpz{A.~Belavin, A.~P.~Polyakov, and
A.~B.~Zamolodchikov, Nucl.~Phys.~{\bf B241}\hfill\break (1984)333}.
The relation to be derived below is a weak generalization of this
remarkable property of conformal field theory.

Let us consider an asymptotic free field theory with one dimensionless
parameter $g$ in the $D$-dimensional euclidean space.  (Examples:
$g$ is the temperature for the O(N) non-linear sigma model in $D=2$, and
the strong fine structure constant for QCD without quarks in $D=4$.)
Let $\beta (g)$ be the beta function of the parameter:
\eqn\ebeta{\dt g = \beta(g) \equiv {1 \over 2}~\beta_1 g^2 + ...~.}
Let $g(t)$ be the solution of the above RG (renormalization group) equation
which satisfies the initial condition $g(t=0) = g$.  With a spatial distance
$r$ we can form an RG invariant $g(\ln r)$.  Note the asymptotic behavior
\eqn\egasymp{g(\ln r) \to {1 \over - {\beta_1 \over 2}~\ln r}\quad
{\rm as}\quad r \to 0~.}

Let $\Phi_a$ be a composite field with scale dimension $x_a + \gamma_a (g)$,
where $\gamma_a (g) \equiv \gamma_{a,1} g + ...$ is the anomalous dimension.
The field satisfies the RG equation
\eqn\eRG{\dt \Phi_a = (x_a + \gamma_a (g)) \Phi_a~,}
where we assume no mixing for simplicity of the discussion.

We denote the trace of the stress-energy tensor by $\H$ which satisfies the
canonical RG equation:
\eqn\eRGH{\dt \H = D \H~.}
We consider the OPE of the trace $\H$ and an arbitrary
composite field $\Phi_a$:
\eqn\eOPEH{\int_{|r|=\ep} \dO (r) \H(r) \Phi_a (0)
\buildrel {\ep \to 0} \over \longrightarrow
\C_a^{~b} (\ep;g) \Phi_b (0)~,}
where $\dO$ is the $D-1$ dimensional angular volume element, and
we have taken the angular average.  We are only interested in
the coefficients for which $x_b \le x_a$.
The RG constrains the coefficients as follows:
\eqn\eC{\C_a^{~b} (\ep;g) = {1 \over \ep^{1+x_a-x_b}} \exp \left[
\int_{g(\ln \ep)}^g dx~{(\gamma_a - \gamma_b)(x) \over \beta (x)}\right]
H_a^{~b} (g(\ln \ep))~.}
We wish to show that the leading behavior of the diagonal element
$H_a^{~a} (g)$ is given by
\eqn\eHleading{H_a^{~a} (g) = {\beta_1 \over 2}~\gamma_{a,1}~g^2
+ {\rm O} (g^3)~,}
where $\gamma_{a,1}$ is the first Taylor coefficient of the anomalous
dimension of $\Phi_a$.  This implies that
\eqn\eCleading{\C_a^{~a} (\ep;g) =
{\gamma_{a,1} \over {\beta_1 \over 2}~\ep \ln^2 \ep} +
{\rm O} \left({1 \over \ep \ln^3 \ep}\right)~.}

To derive eqn.~\eHleading, we first recall that
the volume integral of the trace $\H$ generates a scale transformation
or equivalently an RG transformation \ref\rcj{S.~Coleman and R.~Jackiw,
Ann.~Phys.~{\bf 67}(1971)552}.  Treating the short-distance
singularities carefully, we obtain
\eqn\evar{\eqalign{&\quad\int_{|r-r_k| > \ep_k} d^D r~\vev{(\H (r)
- \vev{\H}_g) \Phi_{a_1} (r_1) ... \Phi_{a_n} (r_n)}_g \cr
&\buildrel {\ep_k \to 0} \over \longrightarrow \quad -
\sum_{k=1}^n \left( x_{a_k}
+ r_{k\mu} {\partial \over \partial r_{k\mu}} \right)
\vev{\Phi_{a_1} (r_1) ... \Phi_{a_n} (r_n)}_g \cr
&\qquad - \sum_{k=1}^n S_{a_k}^{~b} (\ep_k;g) \vev{\Phi_{a_1} (r_1) ...
\Phi_b (r_k) ... \Phi_{a_n} (r_n)}_g~.\cr}}
By differentiating the above asymptotic expansion with
respect to $\ep_k$ we obtain the
relation of the coefficient $S_a^{~b}$ to $\C_a^{~b}$
in \eOPEH:
\eqn\eSC{{\partial \over \partial \ep} S_a^{~b} (\ep;g) =
\C_a^{~b} (\ep;g)~.}
The RG constrains the coefficient $S_a^{~b}$ in the form
\eqn\eS{S_a^{~b} (\ep;g) = {1 \over \ep^{x_a-x_b}}
\exp \left[ \int_{g(\ln \ep)}^g dx~{(\gamma_a - \gamma_b)(x) \over \beta(x)}
\right] \sigma_a^{~b} (g(\ln \ep))~.}
Hence, from eqs.~\eC, \eSC, and \eS, we obtain
\eqn\eHsigma{H_a^{~b} (g) = \beta (g) {d \over dg}~\sigma_a^{~b} (g)
+ (- x_a - \gamma_a (g) + x_b + \gamma_b (g)) \sigma_a^{~b} (g)~.}

We note that under the change of normalization
\eqn\eN{\Phi_a \to N_a (g) \Phi_a~,}
the matrix $\sigma_a^{~b}$ in eqn.~\eS\ changes homogeneously
\eqn\echangesigma{\sigma_a^{~b} (g) \to
N_a (g) \sigma_a^{~b} (g) (N_b (g))^{-1}~.}
Especially the diagonal element $\sigma_a^{~a}$ is independent of the
normalization.  On the other hand, the anomalous dimension changes
inhomogeneously
\eqn\echangegamma{\gamma_a (g) \to \gamma_a (g)
+ \beta (g) \partial_g \ln N_a (g)~.}
If we allow $N_a (g)$ which either vanishes or diverges at $g=0$,
even the first Taylor coefficient $\gamma_{a,1}$ becomes normalization
dependent.

We now choose to normalize $\Phi_a$ so that its two-point function
has a non-vanishing  limit as $g \to 0$:
\eqn\elimit{\vev{\Phi_a (r) \Phi_a (0)}_{g \to 0} = {{\cal N}_a \over
r^{2 x_a}}~,}
where ${\cal N}_a$ is a positive constant.  Under this restriction,
only such $N_a (g)$ which is finite and non-vanishing at $g=0$
is allowed, and the first Taylor coefficient $\gamma_{a,1}$ becomes
independent of normalization.  With this convention, we now argue that
\eqn\egammasigma{\gamma_a (g) = \sigma_a^{~a} (g) + {\rm O} (g^2)~.}
To see this, we study the asymptotic expansion \evar\ perturbatively
for an infinitesimal $g$ for the case $n=2$ and $a_1 = a_2 = a$.
The trace $\H$ is proportional to the beta
function $\beta$ which is of order $g^2$, and we expect the right-hand
side of \evar\ to vanish to order $g^2$.  Because of the RG equation
\eqn\eRGeq{\eqalign{&
- \left(2 x_a + r_\mu {\partial \over \partial r_\mu} \right)
\vev{\Phi_a (r) \Phi_a (0)}_g\cr
&\quad = ( - \beta (g) {\partial \over \partial g} +
2 \gamma_a (g) )
\vev{\Phi_a (r) \Phi_a (0)}_g\cr}}
and eq.~\elimit, we find that the diagonal coefficient $S_a^{~a} (\ep;g)$
must agree with the anomalous dimension $\gamma_a (g)$ to first order
in $g$.  Since
\eqn\ediag{S_a^{~a} (\ep;g) = \sigma_a^{~a} (g(\ln \ep))~,}
we obtain eqn.~\egammasigma.  Finally, eqn.~\eHleading\
follows from eqs.~\eHsigma\ and \egammasigma.

The matrix $\sigma_a^{~b}$ have been calculated for the two-dimensional
O(N) non-linear sigma model \ref\rpaper{H.~Sonoda and W.-C.~Su,
Nucl.~Phys.~{\bf B441}(1995)310}.  Here are three examples: ($\beta_1
= {N-2 \over \pi}$)

\no
(i) The spin field $A \equiv \Phi^I (I=1,...,N)$ normalized by
\eqn\enormspin{\vev{\Phi^I (r) \Phi^J (0)}_{g \to 0} = {1 \over N} \delta^{IJ}}
satisfies
\eqn\egammaspin{\gamma_A (g) \simeq \sigma_A^{~A} (g)
\simeq {N-1 \over 4 \pi}~g~.}

\no
(ii) The first order derivative $B \equiv {1 \over \sqrt{g}} \partial_\mu
\Phi^I$
satisfies
\eqn\enormB{\vev{ {1 \over \sqrt{g}} \partial_\mu \Phi^I (r)
{1 \over \sqrt{g}} \partial_\nu \Phi^J (0) }_{g \to 0}
= \delta^{IJ} {1 \over \pi r^2} {N-1\over 2N}
(\delta_{\mu\nu} - 2 {r_\mu r_\nu \over r^2} )~,}
and we find
\eqn\egammaB{\gamma_B (g) = \gamma_A (g) - {\beta (g) \over 2 g}
\simeq \sigma_B^{~B} (g) \simeq {1 \over 4 \pi}~g~.}

\no
(iii) The second order derivative $C \equiv {1 \over g} \partial^2 \Phi^I$
satisfies
\eqn\enormC{\vev{{1 \over g} \partial^2 \Phi^I (r)
{1 \over g} \partial^2 \Phi^J (0)}_{g \to 0}
= \delta^{IJ} {1 \over \pi^2 r^4} {N-1\over N}~,}
and we find
\eqn\egammaC{\gamma_C (g) = \gamma_A (g) - {\beta (g) \over g}
\simeq \sigma_C^{~C} (g) \simeq {-N+3 \over 4 \pi}~g~.}
In the above examples, the relation \egammasigma, which was missed
in ref.~\rpaper, is verified explicitly.

Before concluding this letter, we consider yet another OPE:
\eqn\eOPEHtilde{\int_{|r|=\ep} \dO (r) {r_\mu r_\nu\over \ep}
\H_{\mu\nu} (r) \Phi_a (0) \buildrel {\ep \to 0} \over \longrightarrow
\tilde{\C}_a^{~b} (\ep;g) \Phi_b (0)~,}
where $\H_{\mu\nu}$ is the stress-energy tensor field
which has no anomalous dimension.  The RG implies
\eqn\eCtilde{\tilde{\C}_a^{~b} (\ep;g) =
{1 \over \ep^{x_a-x_b}} \exp \left[
\int_{g(\ln \ep)}^g dx~{(\gamma_a - \gamma_b)(x) \over \beta (x)}\right]
\tilde{H}_a^{~b} (g(\ln \ep))~.}
Using the conservation law for the stress-energy tensor, we obtain
\eqn\econserv{\partial_\mu ( r_\nu \H_{\mu\nu} ) = \H~.}
This implies
\eqn\eHrel{{\partial \over \partial \ep} \int_{|r|=\ep} \dO(r) {r_\mu r_\nu
\over \ep}
\H^{\mu\nu}(r) \Phi_a (0) =  \int_{|r|=\ep}
\dO (r) \H (r) \Phi_a (0)~.}
Hence,
\eqn\eCCtilde{{\partial \over \partial \ep} \tilde{C}_a^{~b}
(\ep;g) = C_a^{~b} (\ep;g)~.}
Comparing this with eqn.~\eSC, we find that the difference between
$\tilde{C}_a^{~b} (\ep;g)$ and $S_a^{~b} (\ep;g)$
is independent of $\ep$.  Therefore, using the RG constraints
\eS\ and \eCtilde, we obtain
\eqn\eHtilde{\tilde{H}_a^{~b} (g) = \sigma_a^{~b} (g) + k_a \delta_a^{~b}~,}
where $k_a$ is a constant.  Since the constant
$k_a$ is the value of $\tilde{H}_a^{~a} (g)$ at $g=0$, we can resort
to the free theory at $g=0$ for its determination.  Then we find
\eqn\ek{k_a = x_a~,}
where $x_a$ is the na\"{\i}ve scale dimension
of $\Phi_a$.  Thus, the leading behavior
of the diagonal term $\tilde{C}_a^{~a}$ is given by
\eqn\eCtildeleading{\tilde{C}_a^{~a} (\ep;g) = x_a
+ {\gamma_{a,1} \over - {\beta_1 \over 2}~\ln \ep}
+ {\rm O} \left( {1 \over \ln^2 \ep}\right)~.}

In conclusion we have derived the following leading behavior
of the OPE's in asymptotic free field theories:
\eqn\eresults{\eqalign{&\int_{|r|=\ep} \dO (r) \H (r) \Phi_a (0) \to
{\gamma_{a,1} \over {\beta_1 \over 2} ~\ep \ln^2 \ep} \Phi_a (0) +
\rm{off~diagonal~terms}\cr
&\int_{|r|=\ep} \dO (r) {r_\mu r_\nu \over \ep} ~\H^{\mu\nu} (r)
\Phi_a (0) \to \left(x_a + {\gamma_{a,1} \over - {\beta_1 \over 2} \ln
\ep}\right) \Phi_a (0) + \rm{off~diagonal~terms}~,\cr}}
where $\gamma_{a,1}$ is the first Taylor coefficient
of the anomalous dimension of the field
$\Phi_a$ which is normalized so that its two-point function has
a non-vanishing limit at $g=0$ (eqn.~\elimit).
The above OPE's constitute a weak generalization of the structure of
the OPE of the stress-energy tensor and an arbitrary conformal field
in two dimensional conformal field theory.
The extension of eqs.~\eresults\ to asymptotic free field theories
with more than one parameter is straightforward.

Analogous results for the equal-time
commutator of $\H$ and the elementary field $\phi$ in the perturbative
$\phi^4$ theory was obtained earlier
in refs.~\rcj\ and \ref\rwilson{K.~G.~Wilson,
Phys.~Rev.~{\bf D2}(1970)1478}, and more recently in
ref.~\ref\rsonoda{H.~Sonoda,
``The Energy-Momentum Tensor in Field Theory I,'' hep-th/9504113,
unpublished}. Here, we emphasize the importance of
normalizing the composite fields
properly (see \elimit) to get the above results \eresults.

\listrefs
\bye